\documentclass[conference]{IEEEtran}

\usepackage[utf8]{inputenc}
\usepackage[english]{babel}
\usepackage{graphicx}
\usepackage{CJK}
\usepackage[top=2cm, bottom=2cm, left=2cm, right=2cm]{geometry}
\usepackage{algorithm}
\usepackage{algorithmicx}
\usepackage{algpseudocode}
\usepackage{amsmath, amsthm, amsfonts}
\usepackage{tikz}
\usepackage{multirow}
\usepackage{array}
\usepackage{textcomp}
\usepackage{url}

\floatname{algorithm}{}
\renewcommand\IEEEkeywordsname{Keywords}

\begin{document}

\title{A comparative study of neural network techniques for automatic software vulnerability detection }

\author{\IEEEauthorblockN{Gaigai Tang, Lianxiao Meng,\\ Huiqiang Wang*}
\IEEEauthorblockA{School of computer science \\ and technology,\\ Harbin Engineering University,\\ Harbin, China}
\and
\IEEEauthorblockN{Shuangyin Ren, Qiang Wang, Lin Yang*}
\IEEEauthorblockA{National Key Laboratory of Science and \\ Technology on Information System Security, \\ Institute of System Engineering,
 \\ Chinese Academy of Military  Science, \\ Beijing, China}
\and
\IEEEauthorblockN{Weipeng Cao}
\IEEEauthorblockA{College of Computer Science \\ and Software Engineering, \\ Shenzhen University, \\ Shenzhen, China}
}

\maketitle

\begin{abstract}
Software vulnerabilities are usually caused by design flaws or implementation errors, which could be exploited to cause damage to the security of the system. At present, the most commonly used method for detecting software vulnerabilities is static analysis. Most of the related technologies work based on rules or code similarity (source code level) and rely on manually defined vulnerability features. However, these rules and vulnerability features are difficult to be defined and designed accurately, which makes static analysis face many challenges in practical applications. To alleviate this problem, some researchers have proposed to use neural networks that have the ability of automatic feature extraction to improve the intelligence of detection. However, there are many types of neural networks, and different data preprocessing methods will have a significant impact on model performance. It is a great challenge for engineers and researchers to choose a proper neural network and data preprocessing method for a given problem. To solve this problem, we have conducted extensive experiments to test the performance of the two most typical neural networks (i.e., Bi-LSTM and RVFL) with the two most classical data preprocessing methods (i.e., the vector representation and the program symbolization methods) on software vulnerability detection problems and obtained a series of interesting research conclusions, which can provide valuable guidelines for researchers and engineers. Specifically, we found that 1) the training speed of RVFL is always faster than Bi-LSTM, but the prediction accuracy of Bi-LSTM model is higher than RVFL; 2) using {$doc2vec$} for vector representation can make the model have faster training speed and generalization ability than using {$word2vec$}; and 3) multi-level symbolization is helpful to improve the precision of neural network models.

\end{abstract}


\IEEEpeerreviewmaketitle

\section{Introduction} \label{sec-1}
As the functions of modern software systems become more and more complicated, software vulnerabilities caused by design flaws and implementation errors become an inevitable problem in engineering.
According to the statistics released by the Common Vulnerabilities and Exposures (CVE) \cite{cve-info} and  National Vulnerability Database (NVD) \cite{nvd-info}, the number of software vulnerabilities has increased from 1600 to nearly 100000 since 1999 \cite{vul-hack}.
These vulnerabilities pose a serious threat to the security of existing software systems.

Most of the classical vulnerability detection techniques are driven by rules \cite{os-tool-1},\cite{os-tool-2},\cite{os-tool-3},\cite{c-tool-1},\cite{c-tool-2},\cite{c-tool-3} and code similarity metrics \cite{similarity-1},\cite{similarity-2}, which are generated by experts experience.
Although rule-based detection techniques can be applied to vulnerability detection in specific scenarios, the performance of these methods depends heavily on the experience of developers.
Generally, it is very difficult to describe the features of software vulnerabilities accurately, which leads to the corresponding detection rules are also difficult to be defined accurately and completely.

To alleviate the above problems, neural network-based automatic vulnerability detection technology (source code level) has been proposed and shows great potential in related issues \cite{rel-1},\cite{rel-3},\cite{rel-4},\cite{tool},\cite{rel-L-2019}. Neural networks can automatically extract complex features from input data, avoiding the problems of high cost, instability, and incompleteness of manually constructing features and empirically defining rules. A representative work is the VulDeePecker system proposed in \cite{tool}, which is the first effort to use neural network techniques to solve software vulnerability detection problems. Specifically, the VulDeePecker first extracts the $code\ gadget$, which is a form of source code fragments related to the vulnerability, and then symbolizes them to get the corresponding symbolic data. After that,  it uses {$word2vec$} \cite{word2vec-h} to transform the symbolic data into the vector representation and then applies the vectors to the model training of the Bidirectional Long Short-term Memory network (Bi-LSTM) \cite{Bi-LSTM} to get the detection model. The authors have used several experiments to prove that VulDeePecker is superior to other classical vulnerability detection techniques in terms of the accuracy of vulnerability detection.

The success of VulDeePecker shows that it is feasible to use neural networks to detect software vulnerabilities, but there are still some shortcomings in that work, as follows.

\begin{itemize}
\item VulDeePecker only tried the Bi-LSTM with the iterative training mechanism to train the detection model, but this type of neural network often requires a long training time and will face many obstacles in practical applications.
\item VulDeePecker used {$word2vec$} to vectorize the software codes and then used the generated vector (variable length) as the input of the neural network. This method sometimes requires us to perform additional work on the preprocessing of the vector (e.g., padding zeros), which will cause the dimension of the vector to be very high and affect the training efficiency of the model. Moreover, this method may also lose important semantic information about the source codes and affect the effectiveness of the vulnerability detection model.
\item VulDeePecker used two types of  symbolization to do symbolic representation, but it is still unclear which one is the most suitable for neural networks.
\end{itemize}

 The above unsolved problems inspired us to choose two completely different neural networks and different data preprocessing methods to model software vulnerability detection problems in this study. Specifically, we selected the Bi-LSTM with the iterative training mechanism and the Random Vector Functional Link network (RVFL) \cite{RFVL} with the non-iterative training mechanism to train on the vulnerability data set to get corresponding detection models, and then compared the precision and training efficiency of the models. We found that the precision of the Bi-LSTM model is always higher than that of the RVFL model, but the training efficiency of the RVFL is always higher than that of the Bi-LSTM.
Moreover, we have implemented a multi-level symbolization method for symbolic representation and  proposed the use of {$doc2vec$} \cite{doc2vec-h} for vector representation.
In details, we  first perform three symbolizations to obtain the symbolic representations of the source codes  related to the vulnerability.
The advantage of using three levels of symbolization is that one can significantly reduce the noise introduced by irrelevant information in vulnerable codes.
Then we use the {$doc2vec$} to automatically transform the symbolic representations of the source codes to their vector representations.
Our study found that {$doc2vec$} is more suitable for modeling this problem than {$word2vec$} used in \cite{tool}, because it can not only convert source codes with arbitrary length into a fixed-length feature representation, but also better grasp the semantic information. These advantages are helpful to improve the precision and training efficiency of the model. These experimental results can provide valuable guidance for researchers and engineers to choose appropriate neural networks and data preprocessing methods for software vulnerability detection problems.

The rest of this paper is organized as follows.
Section \ref{sec-2} discusses the work related to the automatic detection of software vulnerability.
Section \ref{sec-3} describes the details of the automatic software vulnerability detection system using neural network techniques.
In Section \ref{sec-4}, we give details of our experimental setting, results, and corresponding analysis.
The conclusions and future works are discussed in Section \ref{sec-5}.

\section{Related Work} \label{sec-2}

Traditional machine learning techniques such as Decision Tree \cite{ID3} and Support Vector Machine \cite{SVM},
 mainly extract the vulnerability features from the pre-classified vulnerabilities.
 However, the vulnerability detection patterns generated with this type of features are usually tailored for specific vulnerabilities.
 In \cite{rel-2}, both simple text features (e.g., character count, character diversity, maximum nesting depth
 and complex text features (e.g., character n-grams, word n-grams and suffix trees) are extracted from the source codes and analyzed by using the naive Bayes classifier.
 The experimental results show that simple features performed unexpectedly better compared to the complex features.

Neural network based techniques are able to learn vulnerability features automatically.
 In \cite{tool}, the authors present a vulnerability detection system \emph{VulDeePecker} based on deep learning.
 \emph{VulDeePecker} obtains the samples by first extracting code gadgets (i.e., snippets of codes that are semantically related to the vulnerability)
 from the buggy programs and then transforming into the vector representations.
 The learning algorithm is based on the Long Short-Term Memory (LSTM).
 Experimental evaluations show that \emph{VulDeePecker} outperforms the state-of-the-art vulnerability detection systems in terms of both accuracy and efficiency.
 \cite{rel-3} has implemented various machine learning models for the detection of bugs that can lead to security vulnerabilities in C/C++ code.
 They use features derived from the build process and the source code.
 \cite{rel-4} has developed a vulnerability detection tool based on deep feature representation learning that can directly interpret the parsed source codes.
 The source codes are firstly transformed into tokens and then embedded as vectors for both convolutional neural networks (CNNs) and recurrent neural networks(RNNs).
 Experimental results show that the tool results in high accuracy.
 \cite{rel-5} has proposed a systematic framework for using deep learning to detect vulnerabilities
 that dubbed syntax-based, semantics-based and vector representations (\emph{SySeVR}).
 \emph{SySeVR} can accommodate syntax and semantic information pertinent to vulnerabilities.
 The source codes are successively represented by syntax-based, semantics-based and vector representations.
 Experiments show that \emph{SySeVR} is able to detect several vulnerabilities that are not reported in the National Vulnerability Database.
 \cite{rel-L-2019} performs a quantitative evaluation of the impacts of different factors (e.g., data dependency and control dependency)
 on the effectiveness of neural network based vulnerability detection techniques.
 In \cite{rel-1}, the authors propose a tool called \emph{VuRLE} for automatic detection and repair of vulnerabilities.
 Experiments show that \emph{VuRLE} results in a large improvement on the accuracy of detection and the number of repaired vulnerabilities.

There are also some vulnerability detection techniques that make use of manually-defined features
 \cite{os-tool-1}, \cite{os-tool-2}, \cite{os-tool-3}, \cite{c-tool-1}, \cite{c-tool-2}, \cite{c-tool-3}, or code similarity metrics \cite{similarity-1},\cite{similarity-2}.
 However, there are several drawbacks.
 First, defining vulnerability features is error-prone and costs a lot of manual labor.
 Second, the features can hardly be complete and usually contain only partial information about the vulnerabilites,
 which may lead to a high rate of false-positive and false-negative \cite{tool}.
 Moreover, code similarity based approaches are limited to detecting vulnerabilities caused by code clone.

\section{The Methodology} \label{sec-3}

\begin{figure}
	\centering
	\includegraphics[scale=0.10]{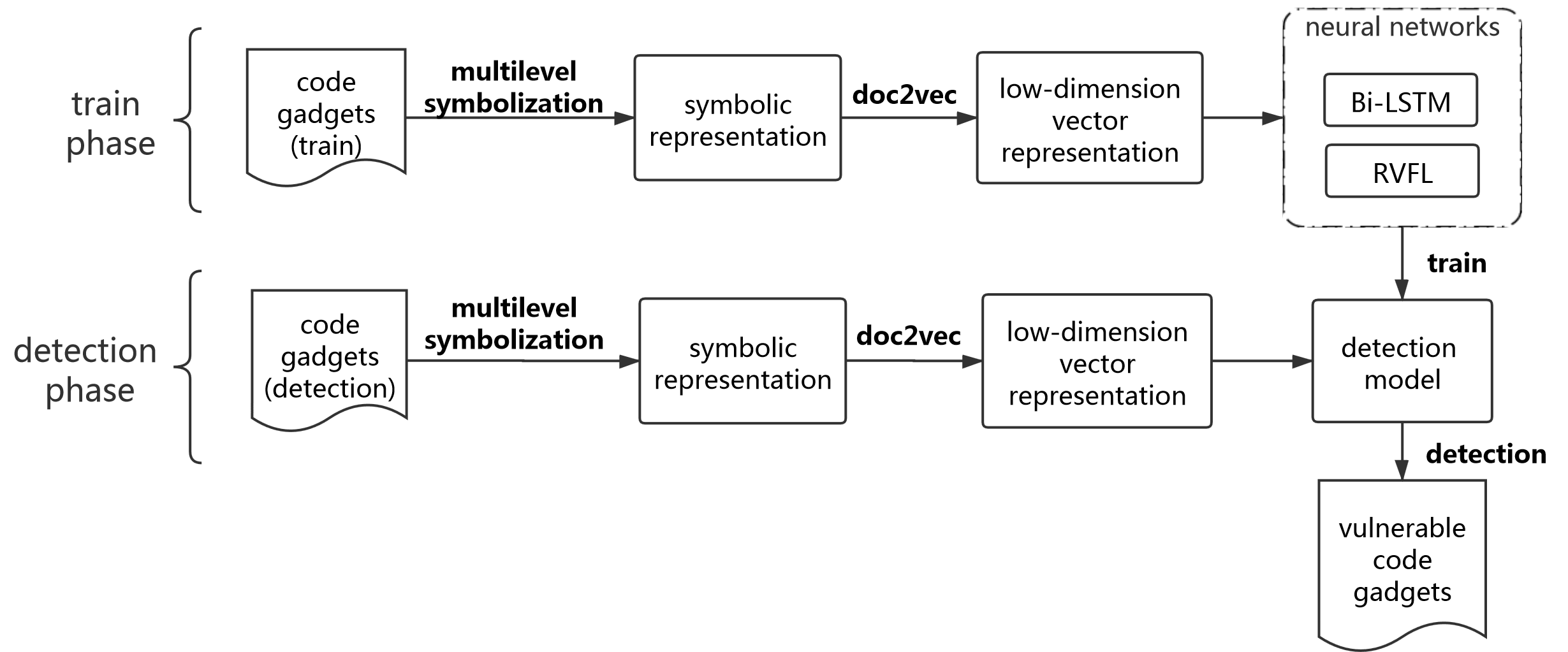}
	\caption{ Overview of the automatic software vulnerability detection system using neural networks}
	\label{fig:overview}
\end{figure}

 We show in Figure. \ref{fig:overview} an overview of the source code level automatic software vulnerability detection system using neural network techniques.
 It first creates a symbolic form of the {$code\ gadget$} by using multilevel symbolization.
 The {$code\ gadgets$} are divided into two parts. One is for training the detection model and the other one is for vulnerability detection.
 Then it transforms the symbolic form into the vector representation with a low dimension by using the tool {$doc2vec$}.
 Finally, it applies two neural network techniques namely the Bi-LSTM and the RVFL to train the detection model.
 In the subsequent sections, we present the details of the main components of the system.

\subsection {Symbolic representation}

 {$Code \  gadget$} is a fragment of the codes that are closely relevant to the vulnerability.
 By using symbolization, {$code\ gadget$} can be further transformed into a symbolic representation.
 The benefit of symbolic representation is that it can effectively reduce the time cost of vector representation precess
 since it can condense the size of the corpus, which is used for training the vector representation tool.
 In symbolization, vulnerability features in each {$code \  gadget$} such as local variables or user-defined functions
 are transformed into certain short and fixed symbolic presentations, where the same features are mapped to the same symbolic presentation.
 In this work, we deploy three symbolization types sorted by priority.

\begin{itemize}
\item Function calls symbolization (F):
 user-defined function names are symbolically represented as FN.
 This symbolization type is set to the first priority because vulnerability is usually caused by improper utilization of library/API function calls.
 Symbolization on user-defined functions can improve the Signal-Noise Ratio (SNR) of the library function in vulnerability information.

\item Variable symbolization (V):
 variable names including parameters and local variables are symbolically represented as VN.

\item Data type symbolization (D): data types of variable and user-defined function are symbolically represented as TN.
 It has the least priority since usually many data types are not related to vulnerability information.
\end{itemize}

The symbol $N$ mentioned above in symbolization is a number representing the sequence of the first occurrence of the feature.
 Besides, all the symbolization types will  reserve the keywords of C/C++ language.

\begin{table}[htbp]
  \centering
  \caption{ multilevel symbolization }
  \label{tab:symbolization}
	\begin{tabular}{|c|c|} \hline
      Symbolization level			& Symbolization group	\\   \hline		
			{Level 1}			     &	F 					\\	 \hline
			{Level 2}			     &	F+V;F+D					\\	 \hline
			{Level 3}			     &	F+V+D				     \\	 \hline
    \end{tabular}
\end{table}

We build a multilevel symbolization mechanism according to the priority of a single symbolization shown in Table. \ref{tab:symbolization}.
 Level 2 includes two symbolization groups F+V and F+D.
 This is because symbolization V and D may have different effects on improving the SNR of vulnerability information with different data-sets.

We take Sample 0 below as an example and explain how the symbolization works.
 We choose the symbolization group F+V in Level 2.
 From Sample 0 we can observe that there are 2 user-defined functions, 5 variables and 2 data types.

 \begin{itemize}
 \item In Level 1, the two user-defined functions are symbolically represented as F1 and F2.
 \item In Level 2, the five variable names are symbolically represented as $Vi, i\in [1,5]$.
 \item In Level 3, the two data types are symbolically represented as T1 and T2.
  \end{itemize}

As a result, with the three levels of symbolization,
 Sample 0 is gradually simplified to a generalized symbolic representation, which may effectively characterize different manifestations of the same vulnerability.

	\begin{CJK*}{UTF8}{gkai}
    \begin{algorithm}
				Sample 0
        \begin{algorithmic}[1] 
				\State {$static \  void \  goodG2B()$}
				\State {$list < char * >\  dataList ;$}
			  \State {$goodG2BSink ( dataList ) ;$}
				\State {$void \  goodG2BSink(list<char *>\  List) $}
				\State {$char * data = List.back ();$}
				\State {$if ( sscanf ( data , " \&d" , \& n ) == 1 ))$}
        \end{algorithmic}
    \end{algorithm}
\end{CJK*}

\begin{CJK*}{UTF8}{gkai}
    \begin{algorithm}
				Level 1 F
        \begin{algorithmic}[1] 
        \State{$static\  void\  F1()$}
				\State{$list < char * >\  dataList ;$}
				\State{$F2 ( dataList );$}
				\State{$void \  F2(list<char *>\  List) $}
				\State{$char * data = List . back ( );  $}
				\State{$if ( sscanf (  data ," \&d" , \& n) == 1 ))$}
        \end{algorithmic}
    \end{algorithm}
\end{CJK*}

\begin{CJK*}{UTF8}{gkai}
    \begin{algorithm}
				Level 2 F+V
        \begin{algorithmic}[1] 
        \State{$static\  void\  F1()$}
				\State{$list < char * >\  V1 ;$}
				\State{$F2 ( V1 );$}
				\State{$void F2(list<char *>\  V2) $}
				\State{$char * V3 = V2 . back ( );  $}
				\State{$if ( sscanf (  V3 , V4 , \& V5 ) == 1 ))$}
        \end{algorithmic}
    \end{algorithm}
\end{CJK*}

\begin{CJK*}{UTF8}{gkai}
    \begin{algorithm}
				Level 3 F+V+D
       \begin{algorithmic}[1] 
       \State{$T1\  F1()$}
				\State{$list < T2 * >\  V1 ;$}
				\State{$F2 ( V1 );$}
				\State{$void\  F2(list<T2 *>\  V2) $}
				\State{$T2 * V3 = V2 . back ( );  $}
				\State{$if ( sscanf (  V3 , V4 , \& V5 ) == 1 ))$}
        \end{algorithmic}
    \end{algorithm}
\end{CJK*}

\subsection {Vector representation}
Since the input type accepted by neural networks is a vector, it still needs to be converted to the vector representation after symbolizing the source codes.
Currently, the most popular vectorization methods are {$word2vec$} \cite{word2vec} and {$doc2vec$} \cite{doc2vec}.

{$word2vec$} is a distributed representation method for words.
 Compared with the one-hot representation, a high-dimensional and sparse representation method, $word2vec$ can get low-dimensional and dense vector representation, which is conducive to the improvement of training efficiency and accuracy of the model, so it has been widely used in vulnerability detection scenarios recently \cite{rel-3}, \cite{tool}, \cite{rel-L-2019}. However, $word2vec$ is not suitable for vectorizing sentences or documents because it ignores the influence of the word order on the information of sentence or document.

 $doc2vec$ was proposed in \cite{doc2vec}, where the authors proposed the unsupervised algorithm {$Paragraph\ Vector$} that can learn fixed-length feature representations from variable-length pieces of texts, ranging from sentences to documents. The $Paragraph \ Vector$ can memorize the missing content in the current context or the topic of the paragraph, so it can extract global features better than $word2vec$.

The input vector of neural networks is required to be fixed length.
{$word2vec$} converts words to vector representations in a one-to-one fashion, therefore the length of the converted vector representation varies with the length of the text.
To satisfy the requirement of neural networks, the vector representations generated by {$word2vec$} need to be further processed to get the corresponding fixed-length version.
Different from $word2vec$, $\ doc2vec$ can directly get fixed-length vector representations from pieces of texts with arbitrary length. As mentioned above, $\ doc2vec$ can also extract more semantic information from the context of the text than $word2vec$.
Therefore, one can infer that $\ doc2vec$ is more suitable than $word2vec$ in the scenarios with rich semantic information.

\subsection {Neural network model}
Here we introduce two types of neural networks, that is, Recurrent Neural Network (RNN) \cite{RNN} using the iterative training mechanism and RVFL \cite{RVFL-1} using the non-iterative training mechanism.

\begin{figure}
	\centering
	\includegraphics[scale=0.3]{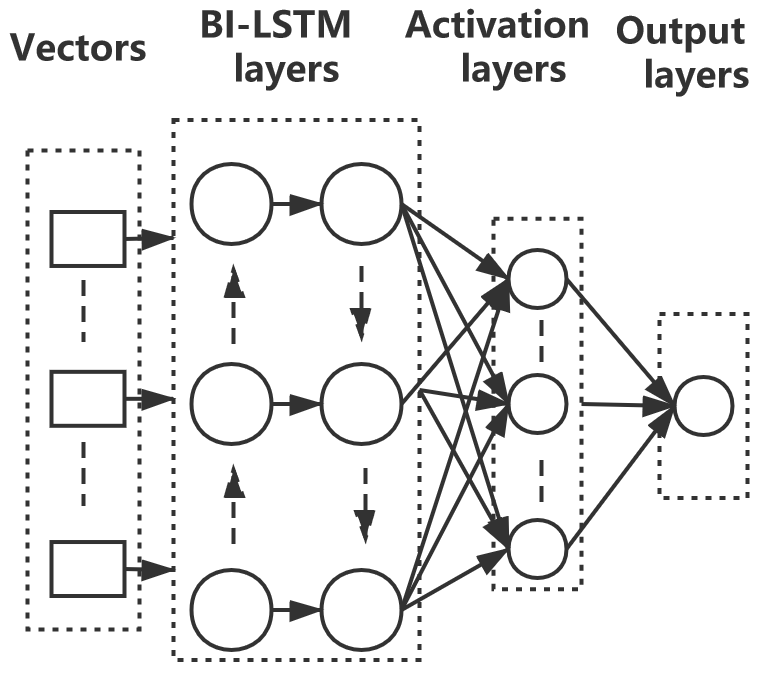}
	\caption{ A typical network structure of Bi-LSTM }
	\label{fig:Bi-LSTM}
\end{figure}

RNN is a typical neural network for processing sequence data \cite{RNN}\cite{BiRNN}. One of its biggest advantages is the ability to mine the context information of the data.
However, since the original RNN can only mine the text information in a forward way, while the vulnerability information usually relates to both the preceding and backward parts of the codes, it can be inferred that the original RNN may not be suitable for dealing with the vulnerability detection problems.
For this reason, we choose the improved version of RNN, that is, Bi-LSTM, which can not only express the long-term dependency information in the input data but also can bi-directionally mine the context information of the data (i.e., forward and backward).
These advantages make Bi-LSTM very suitable for processing software vulnerability detection data (source code level). Note that all parameters of the Bi-LSTM are iteratively fine-tuned using methods like gradient descent. A typical network structure of Bi-LSTM is shown in Figure. \ref{fig:Bi-LSTM}.

\begin{figure}
	\centering
	\includegraphics[scale=0.25]{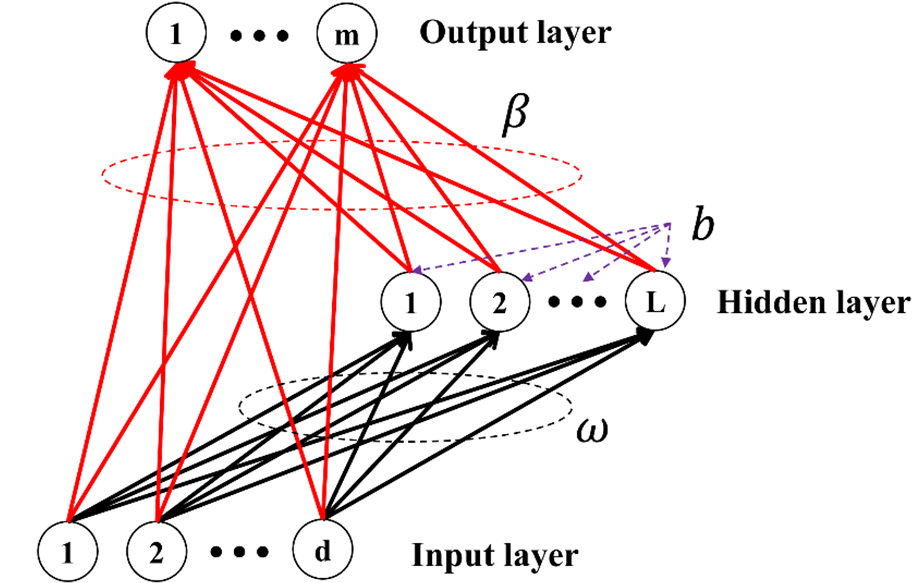}
	\caption{ A typical RVFL with a single hidden layer network structure}
	\label{fig:RVFL}
\end{figure}

Different from Bi-LSTM, RVFL is a special type of feed-forward neural networks with the non-iterative training mechanism.
RVFL was proposed by Pao YH et al. in the 1990s \cite{RVFL-1}, which randomly assigns values to some parameters according to certain rules and keeps these parameters be frozen throughout the training process,
 while other parameters are calculated by the least square method.
This training mechanism can bring RVFL much faster training speed than traditional neural networks with the iterative training mechanism on some tasks with a relatively small data scale.
 In recent years, RVFL and its variants have been widely concerned and applied in many scenarios \cite{RVFL-2}.
 The network structure of the RVFL is shown in Figure. \ref{fig:RVFL}. For the technical details of RVFL, please refer to \cite{RVFL-3}.

\section{Experiment and Evaluation} \label{sec-4}

Our aim is to quantitatively evaluate the performances of the neural network based automatic software vulnerability detection techniques.
  To be specific, we investigate the following questions in the experiments.

\begin{itemize}
\item Question1 (Q1):
How different neural network models perform on vulnerability detection on the source code level?
\item Question2 (Q2):
How different vector representation methods affect the performances of neural networks?
 Specifically, does {$doc2vec$} outperform {$word2vec$} for the task of vulnerability detection?
\item Question3 (Q3):
 What are the effects of different symbolization on the performances of neural networks?
\end{itemize}

\subsection{Experiment setting and implementation}

\subsubsection{Data-set}

We include the following three data-sets from \cite{tool} in our experiments.
 Each sample is a source code file with known vulnerabilities.
 In Table. \ref{tab:dataset} we show the number of samples (i.e., source code files) in each data-set.

\begin{itemize}
\item BE-ALL: samples with Buffer Error vulnerabilities (CWE-119) and ALL library/API function calls.
\item RM-ALL: samples with Resource Management error vulnerabilities (CWE-399) and ALL library/API function calls.
\item HY-ALL: sample with HYbrid  buffer error vulnerabilities (CWE-119) and resource management error vulnerabilities (CWE-399) and ALL library/API function calls.
\end{itemize}

\begin{table}[htbp]
  \centering
  \caption{ number of samples in each data-set }
  \label{tab:dataset}
	\begin{tabular}{|c|c|p{45pt}<{\centering}|p{55pt}<{\centering}|} \hline
      Data-set			& Code Gadgets  & Vulnerable code gadgets	&	Not vulnerable code gadgets	\\   \hline		
			{BE-ALL}			&	39753 			&	10440					&	29313					\\	 \hline
			{RM-ALL}			&	21885				&	7285					&	14600					\\	 \hline
			{HY-ALL}			&	61638				&	17725					&	43913					\\	 \hline
    \end{tabular}
\end{table}

Each data-set is partitioned into two parts with a proportion of 80\% and 20\%,
 where the larger part is for training and the other one is for detection.
 All samples in the data-set are in the form of code gadgets with ground truth labels.

\subsubsection{Parameters setting for neural networks}

In our experiments, we used two types of neural networks for model training: Bi-LSTM and RVFL.
 For both, there is only one hidden layer in the network structure.
 In Bi-LSTM, hidden layer neurons perform two-time computation operations (i.e., forward and backward),
 while in RVFL they only perform a one-time computation operation.
 We considered a one-time computation operation of the neurons as a unit of measurement. To make a fair comparison, we set the number of hidden layer neurons of RVFL to twice that of Bi-LSTM, which guarantees the structural complexity of the two neural networks is essentially the same. We have implemented the CPU versions of Bi-LSTM and RVFL, and all the models are trained in the PC environment with CPU. For Bi-LSTM, the batch size, dropout rate, the number of epochs, and the number of the hidden layer neurons were set to 64, 0.5, 2, and 60, respectively, and the RMSProp was chosen as the optimizer. For RVFL, the number of the hidden layer neurons was set to 120 and the Sigmoid function was chosen as the activation function. The input weights and the hidden biases of RVFL were generated randomly from (-1, 1) and (0, 1) under a uniform distribution, respectively.

\subsection {Results and evaluation}

\subsubsection {Evaluation metrics}
In our experiment, we used the same indexes as \cite{evaluation} to evaluate the effectiveness of the vulnerability detection model, that is, False Positive Rate (FPR), False Negative Rate (FNR), Precision (P), and F1-measure (F1).
The value range of these four indicators is [0, 1]. For FPR and FNR, the closer their values are to 0, the better the performance of the model; for other indicators, the closer their values are to 1, the better the performance of the model.

The quality of vector representation can be evaluated by {$CosineDistance$} ({$cosine$}) between vectors in the vector space, which can be calculated by the following formula.
The range of the $cosine$ value is [0,1]. The closer the value is to 1, the more similar the two vectors are.

\begin{equation*}
cosine(A,B)=\frac{A \cdot B}{||A||_2||B||_2}
\end{equation*}
where A and B refer to vectors.


In our experiments, we have the following three configurations:

\begin{itemize}
\item  {$word2vec$} with Bi-LSTM (w+B), which was used by VulDeePecker;
\item {$doc2vec$} with Bi-LSTM (d+B);
\item {$doc2vec$} with RVFL (d+R).
\end{itemize}

\subsubsection {Results for Q1}

Regarding the impacts of different neural network models on the performances of vulnerability detection.
 We evaluate the precision of the above three configurations on all data-sets,
 while we evaluate the efficiency of configuration d+B and d+R on the largest data-set HY-ALL.
 In the experiments, all the three data-sets are preprocessed with the symbolization group F+V.

 \begin{table}[htbp]
  \centering
  \caption{ effect of different vector representation tools  on vulnerability detection precision}
  \label{tab:Effectiveness}
	\begin{tabular}{|p{32pt}<{\centering}|c|c|c|c|c|} \hline
     Data-set  	&Pattern 				& FPR(\%)			 & FNR(\%)  	&	P(\%) 	& F1(\%)	\\ \hline		
\multirow{3}{*}{BE-ALL}	&w+B   &2.9                    &18.0				&91.7			&	86.6		\\	
        \cline{2-6}	&d+B	&3.9	 				&16.9				&88.1			&	85.5	\\
        \cline{2-6}	&d+R		 &6.7	 			    &32.5				&85.1			&	74.3	\\	\hline
\multirow{3}{*}{RM-ALL} &w+B   &2.8	 				&4.7				&94.6			&	95.0		\\	
       	\cline{2-6}	&d+B	&3.8	 				&9.7				&91.9			&	91.1\\	
        \cline{2-6}	&d+R		&3.1	 				&20.0				&90.5			&	86.5	\\	 \hline
\multirow{3}{*}{HY-ALL} &w+B	&5.1	 				&16.1				&86.9			&	85.4		\\
        \cline{2-6}	&d+B	&3.3					&16.2				&91.0 		    &87.2		\\	
        \cline{2-6}	&d+R		&7.7	 				&29.7				&85.0			&	75.6	\\	 	\hline
    \end{tabular}
\end{table}

As illustrated in Table. \ref{tab:Effectiveness}, the configurations d+B and w+B outperform d+R on the metrics P and F1 for all the three data-sets.
In terms of FPR and FNR, the configurations w+B and d+B have a smaller sum of FPR and FNR than d+R.
The general result is that the configurations with Bi-LSTM perform better than the ones with RFVL on the precision of vulnerability detection.
This can be explained by the fact that Bi-LSTM is able to express the long-term dependency information in the input,
while RVFL is based on forward neural network, which is slightly inferior to Bi-LSTM in context processing.

\begin{table}[htbp]
  \centering
  \caption{effect of different neural network models on vulnerability detection efficiency}
  \label{tab:Complexity}
	\begin{tabular}{|c|p{45pt}<{\centering}|p{45pt}<{\centering}|p{25pt}<{\centering}|p{27pt}<{\centering}|} \hline
      Pattern 					& Training code gadgets		 & Detection  code gadgets  	&	Training Time(s) 	& Detection  Time(s)	\\   \hline
			w+B &	48744	 	&12894	&36372.2		&	156.2		\\	 \hline
			d+B					 &	49310	  &12328	&1123.5				&3.2		\\	 \hline
            d+R				 &	49310	  &12328	&4.62				&0.16		\\	 \hline
    \end{tabular}
\end{table}

From Table. \ref{tab:Complexity},
we can observe that the configuration w+B takes the longest time for training and detection on the data-set HY-ALL,
 while the configuration d+B takes less than 1/30 of the time needed by w+B.
 It's because that configuration w+B in \cite{tool} outputs vectors with a long dimension of 2500,
 which creates a higher computation complexity for the neural networks.
 Moreover, compared with the configuration d+B, configuration d+R further improves the efficiency of training and detection to the level of a few seconds.
 This can be explained by the fact that the non-iterative training mechanism of RVFL reduces the computation of parameters, however with a sacrifice of precision.

We can conclude that the configuration with Bi-LSTM achieves a higher precision, while the configuration with RVFL is more effective.
 Bi-LSTM results in  a high accuracy for detecting vulnerabilities because of its advantage of context processing,
 while RVFL results in a high speed for training model because of its non-iterative training mechanism.
 In the practice, we usually need to make a trade-off between the prediction accuracy and efficiency of the model.
 Specifically, if the speed of model training takes precedence over its accuracy, then RVFL is recommended; otherwise, Bi-LSTM is recommended.

\subsubsection {Results for Q2}

To answer the second question, we evaluate the accuracy of  the two vector representation methods {$doc2vec$} and {$word2vec$}.
 We implement experiments with the samples shown in Sample 1, Sample 2 , Sample 3 and Sample 4.
 Sample 2 and Sample 4 are labeled as 'vulnerable' while the Sample 1 and Sample 3 are not.
 All samples are from the data-set BE-ALL and HY-ALL.
 We evaluate vector representations by using the similarity measure {$cosine$}.
 Both samples are preprocessed with symbolization group of F+V.
 The vector dimension of  {$word2vec$} is set to be 2500, where the vector dimension of one word is set to 50 and the number of words to represent a paragraph is set to be 50.
 The vector dimension of  {$doc2vec$} is set to be 250.
 The different dimension settings of {$word2vec$} and {$doc2vec$} is because that if both dimensions are set to be the same (e.g. 250),
 then for {$word2vec$} the vector dimension of one word will be 5, or the number of words to represent a paragraph will be 5,
 which may have a great impact on the effectiveness of vector representation.
 As a result, the comparison of the accuracy of {$word2vec$} and {$doc2vec$} is carried out under the condition that they both use a proper dimension of vector representation.

\begin{CJK*}{UTF8}{gkai}
    \begin{algorithm}
				Sample 1
       \begin{algorithmic}[1] 
				\State{$data = (char *)malloc(100*sizeof(char)); $}
				\State{$goodG2BSource(data);$}
                \State{$void \ goodG2BSource(char * \&data)$}
                \State{$memset(data, 'A', 50-1);$}
                \State{$data[50-1] = '\backslash 0';$}
                \State{$char\  dest[50] = "";$}
                \State{$strcpy(dest, data);$}
        \end{algorithmic}
    \end{algorithm}
\end{CJK*}

\begin{CJK*}{UTF8}{gkai}
    \begin{algorithm}
				Sample 2
       \begin{algorithmic}[1] 
                \State{$char * ~~ data;$}
			    \State{$data = (char *)malloc(100*sizeof(char));$}
				\State{$if(5==5)$}
                \State{$memset(data, 'A', 100-1);$}
                \State{$data[100-1] = '\backslash 0';$}
                \State{$char \  dest[50] = "";$}
                \State{$strcpy(dest, data);$}
        \end{algorithmic}
    \end{algorithm}
\end{CJK*}

\begin{CJK*}{UTF8}{gkai}
    \begin{algorithm}
				Sample 3
       \begin{algorithmic}[1] 
                \State{$data = -1;$}
			    \State{$char \ inputBuffer[CHAR\_ARRAY\_SIZE] ='''';$}
                \State{$if (fgets(inputBuffer, CHAR\_ARRAY\_SIZE, $}
                \Statex{$ stdin)!= NULL)$}
        \end{algorithmic}
    \end{algorithm}
\end{CJK*}

\begin{CJK*}{UTF8}{gkai}
    \begin{algorithm}
				Sample 4
       \begin{algorithmic}[1] 
			    \State{$char \ inputBuffer[CHAR\_ARRAY\_SIZE] ='''';$}
                \State{$if (fgets(inputBuffer, CHAR\_ARRAY\_SIZE, $}
                \Statex{$ stdin)!= NULL)$}
                \State{$data = atoi(inputBuffer);$}
        \end{algorithmic}
    \end{algorithm}
\end{CJK*}

\begin{table}[htbp]
  \centering
  \caption{effectiveness of word2vec compare with doc2vec}
  \label{tab:Effectiveness-1}
	\begin{tabular}{|c|c|p{42pt}<{\centering}|p{42pt}<{\centering}|p{42pt}<{\centering}|} \hline
    Sample  &Tool 					& Vector dimension		 & Cosine(BE-ALL)  	  & Cosine(HY-ALL) 	\\   \hline
	\multirow{2}{*}	{1\&2}	&word2vec			&	2500	 		 &0.832		       &0.828		\\	
		          \cline{2-5}     	&doc2vec				&	250	 			 &0.642		       &0.639		\\	 \hline
    \multirow{2}{*}   {3\&4}   &word2vec			&	2500	 		 &0.482		       &0.514		\\
                    \cline{2-5}    &doc2vec			&	250	 		 &0.586		       &0.578		\\            \hline
    \end{tabular}
\end{table}

From Table. \ref{tab:Effectiveness-1},
we observe that for Sample1 and Sample2 {$word2vec$} outputs vectors with a higher {$cosine$} than {$doc2vec$} ,
 while for Sample3 and Sample4 it outputs a opposite outcome.
 Obviously, the similarity between  Sample1 and Sample2 is not as high as the result of {$word2vec$}, but it is close to the result of {$doc2vec$}.
 While the similarity between Sample3 and Sample4 is closer to the result of {$doc2vec$},
 proving that {$doc2vec$} can generate a more accurate vector representation with lower dimension than {$word2vec$}.
 It can be explained by the fact that as noted in \cite{tool},
 in order to obtain a fixed length of vector representation,
 vectors generated from {$word2vec$} should be padded with zeros, which may cause the loss of semantic information of the samples.
 Moreover, it is understood that vectors with low dimensions can speed up the training process of the neural network model.
 This is also justified by the results in Table. \ref{tab:Effectiveness},
 where the configurations with $doc2vec$ show better results than the ones with $word2vec$ due to the benefits gained from the vector representation of $doc2vec$.

\subsubsection {Results for Q3}

Given that fact that the configuration d + B has the highest precision of vulnerability detection,
 we take the configuration d + B as the baseline to discuss whether symbolization can further improve the accuracy.
 We implement experiments with all the three data-sets.
 For each data-set, we apply the symbolization level from 1 to 3 for preprocessing.

\begin{table}[htbp]
  \centering
  \caption{ Effectiveness of symbolization levels on precision}
  \label{tab:Effectiveness-2}
	\begin{tabular}{|c|p{45pt}<{\centering}|c|c|c|c|} \hline
         Data-set   	& Symbolization	group		& FPR(\%)		 & FNR(\%)  	&	P(\%) 	& F1(\%)	\\   \hline
 \multirow{4}{*} {BE-ALL} 		&	F	&	2.9	&	13.8	&91.2 &88.6		\\	
		 \cline{2-6}		&	F+V			&	3.9	&	16.9	&	88.1 &85.5		\\	
         \cline{2-6}		&	F+D			&	3.2	&	15.4	&	90.4 &87.4		\\
		 \cline{2-6}		&	F+V+D		&	3.5	&15.5		&89.4 	&86.9		\\	 \hline
 \multirow{4}{*} {RM-ALL} 		&	F	&	2.8	&7.8	&	94.1 &93.1		\\	
		 \cline{2-6}		&F+V			&	3.8	&9.7	&	91.9 &91.1		\\	
         \cline{2-6}		&	F+D			&	2.6	&6.5	&	94.5&94.0		\\
		 \cline{2-6}		&	F+V+D		&	3.4	&9.1		&92.7 	&91.8		\\	 \hline
\multirow{4}{*}{HY-ALL}			&	F    &3.2 &	12.2	&	92.0 &89.8		\\	
		 \cline{2-6}		&F+V			&	3.3	&	16.2	&	91.0 &87.2		\\	
         \cline{2-6}		&	F+D			&3.2	&	14.4	&	91.7 &88.5		\\
		 \cline{2-6}		&	F+V+D		&	3.3	&14.0		&91.1 	&88.5		\\	 \hline
    \end{tabular}
\end{table}

Table. \ref{tab:Effectiveness-2} summarizes the results of how different levels of symbolization can affect the precision.

  From the perspective of data-set types, different symbolization levels have a bigger impact on the precision of vulnerability detection with smaller data-sets, which show the maximum precision deviation of 3.1\% in BE-ALL and  2.6\% in RM-ALL. However, in large data-set HY-ALL, this phenomenon is much weakened, which is shown as 0.9\%. This may be because since the richness of data-sets improves the generalization ability of the detection model itself, and the impact of symbolization is gradually reduced.
  From the perspective of symbolization levels, the configuration d + B with the symbolization level of 1 shows a better and more stable performance than other symbolization levels, while the symbolization level of 2 results an unstable performance, and the symbolization level of 3 outcomes the worst performance. The main reason is that with a high level of symbolization we may lose some key vulnerability information in the source codes. Moreover, it should be mentioned that symbolization groups of F +D  outperforms than symbolization groups of  F + V with all data-sets, it may due to there are many contents related to data type in the source codes,  after symbolizing them can better capture the vulnerability information.

\section{Conclusions} \label{sec-5}

In this paper, we conducted a comprehensive experiment to analyze the impact of the types of neural networks and different data preprocessing methods on solving software vulnerability detection problems.
Our experimental results show that the RVFL outperforms Bi-LSTM on the efficiency of vulnerability detection, while Bi-LSTM performs better on the precision.
Moreover, vector representation using {$doc2vec$} and an appropriate level of symbolization can effectively improve the accuracy of vulnerability detection.
These experimental conclusions will provide researchers and engineers with guidelines when choosing neural networks and data preprocessing methods for vulnerability detection.

In the future, we will study the combination of Bi LSTM and RVFL to design a new neural network and use it for source code level vulnerability detection.
It is expected that the new method will meet real-life engineering needs in terms of efficiency and precision.

\section*{Acknowledgment}

This research is partially supported by the vulnerability analysis technology for UAV communication protocol project
 of National Key Laboratory of Science and Technology on Information System Security, the Natural Science Foundation of China (No.~ 61872104), and the Opening Project of Shanghai Trusted Industrial Control Platform
(TICPSH202003008-ZC).

\bibliographystyle{unsrt}
\bibliography{TASE}

\end{document}